\documentclass[prb,twocolumn, superscriptaddress,showpacs, showkeys]{revtex4}

\usepackage{graphicx}
\usepackage{color}

\begin{document}

\title{Structural and magnetic transition in CeFeAsO: separated or connected}

\author{A. Jesche}
\author{C.\,Krellner}
\affiliation{Max Planck Institute for Chemical Physics of Solids, D-01187 Dresden, Germany}
\author{M. de Souza}
\author{M. Lang}
\affiliation{Physikalisches Institut, Goethe-Universit\"at Frankfurt, D-60438 Frankfurt(M), Germany}
\author{C.\,Geibel}
\affiliation{Max Planck Institute for Chemical Physics of Solids, D-01187 Dresden, Germany}
\email{jesche@cpfs.mpg.de}

\date{\today}

\begin{abstract}
Using an adapted Sn-flux growth technique we obtained comparatively large CeFeAsO single crystals of better quality than previously reported polycrystals or single crystals, as evidenced by much sharper anomalies at the structural and magnetic phase transitions as well as a much higher residual resistivity ratio of 12. 
In the magnetically ordered phase we observe a very pronounced metallic behavior of the in-plane resistivity, which excludes a Mott insulator regime at low temperature. 
The separation $\Delta T = T_0 - T_N$ between structural and magnetic ordering temperatures decreases with increasing sample quality, from 18\,K in the initial reports to 6\,K in the present single crystals, demonstrating that this separation is not an intrinsic property of the RFeAsO systems. 
Our results indicate that the coupling between magnetic ordering and structural distortion is very similar in AFe$_2$As$_2$ and RFeAsO type of compounds, much more similar than previously thought. 
The implications of our experimental results give arguments both in favor and against the nematic phase model.
\end{abstract}



\maketitle
\section{Introduction}
The discovery of superconductivity with $T_c = 26$\,K in fluorine doped LaFeAsO \cite{kamihara2008} caused enormous interest in this class of materials and led to the discovery of superconductivity in several other layered iron-pnictide compounds.\cite{ishida09} 
A common feature of most of these systems is a structural and a magnetic transition at temperatures $\approx 150 - 220$\,K in the undoped materials.\cite{delacruz2008} 
Doping or application of pressure lead to the suppression of the structural and the magnetic transitions and to the appearance of superconductivity. 
While the whole research field started with the RFeAsO (1111) compounds (R: rare earth)\,\cite{takahashi08,ren08a,chen2008}, later on the focus shifted towards the AFe$_2$As$_2$ (122) systems (A = Ba\,\cite{rotter08, sefat08, ni08_2}, Sr\,\cite{krellner2008,chen08-122,sasmal08,yan08}, Ca\,\cite{ronning08,ni08_3}, Eu\,\cite{jeevan08,ren08-122}), despite their lower $T_c$.\cite{ishida09}.  
The reason was a simple but essential material-related problem; sample preparation and especially single crystal growth is easier for the 122 than for the 1111. 
Thus, comparatively large, high-quality 122 single crystals have been available for at least one year, allowing for many investigations which could not be carried out on the 1111 systems, because size and/or quality of 1111 single crystal is still limited. 
As a result, the physical properties of the 122 compounds are presently much better known than those of the 1111. 
However, for some aspects the 1111 systems seem to be the more interesting ones. 
They still present the highest $T_c$ among the Fe-based superconductors, being only surpassed by the cuprates. 
This is likely to be related to a weaker bonding and exchange along the $c$-axis resulting in a stronger two-dimensional character.\cite{qi09} 
The 1111 compounds also present some distinct differences to the 122, which are quite relevant for a fundamental understanding of these systems. 
While a well-defined metallic state is established for the undoped 122 compounds, resistivity results on undoped LaFeAsO polycrystals\,\cite{kamihara2008} and single crystals\,\cite{yan09-arxiv,chen09}  systematically show an increase towards low temperatures, suggesting the proximity to a metal-insulator transition. 
The closeness to such a transition and the related question on the strength of correlation effects in these systems are key problems in the field.\cite{si09} 
A second, intriguing difference concerns the relation between the magnetic and the structural transition in the undoped systems. 
All present experimental results on the 1111 indicate that the structural transition from the tetragonal to the orthorhombic structure occurs first upon cooling, while the long-range antiferromagnetic order sets in at a slightly lower temperature, 10 - 20 K below.\cite{delacruz2008,zhao2008,luo09} 
In contrast, for the 122 compounds it was clear from the first experiments that both transitions occur simultaneously\,\cite{jesche08}, and it is meanwhile well established that the common transition is a first-order type one. 
This experimentally observed difference between 1111 and 122 is in clear disagreement with theoretical results from LDA-based \textit{ab initio} calculations, which imply an intimate connection between the two transitions in both type of compounds.\cite{krellner2008}
Since the relation between structural and magnetic transition is also a central issue towards a deeper understanding of these systems, it is crucial to find out the origin for this difference.
Over the past few years we have developed a high temperature Sn-flux technique for the growth of RTPO (T: transition metal) single crystals.\cite{krellner08Ru} We recently adapted this method to CeFeAsO, and obtained larger single crystals than those obtained from high pressure techniques\,\cite{zhigadlo08}, and of much better quality than those recently reported from NaAs flux growth.\cite{yan09,yan09-arxiv} 

Here we present a study of the 3$d$-related physics using resistivity, $\rho(T)$, specific heat, $C(T)$, and thermal expansion measurements, $\alpha(T)$, with emphasis placed on the structural and the magnetic transition. The $4f$-related physics at lower temperatures was already presented and discussed in a recent publication.\cite{jesche09} Our results show a well-defined metallical behavior in the basal plane within the magnetic ordered state, resulting in a residual resistivity ratio (RRR) of unprecedented $\approx 12$. 
The structural and the magnetic transitions are much sharper in our single crystals than in previously reported poly- or single crystals, and the separation between $T_0$ and $T_N$ decreases with increasing sample quality, down to 6 K compared to the initially reported 20 K. 
We discuss the implication for the relation between magnetic and structural degrees of freedom.

\section{Experimental}
The samples were synthesized using a two-step Sn-flux technique. 
In a first step, As and Sn were heated up to 600$^{\circ}$C for 5\,h in an alumina crucible which was sealed inside an evacuated silica ampule. 
In a second step, Ce, Fe, SnO$_2$, and Sn were added and the alumina crucible was sealed inside a Ta-container under argon atmosphere. The constituents were mixed in a molar ratio of Ce:Fe:As:O:Sn = 2:1:2:2:5. 
The mixture was then heated up to 1500$^{\circ}$C, slowly cooled down to 900$^{\circ}$C within one week followed by fast cooling down to room temperature (RT). 
To remove the excess Sn, the samples were centrifugated at 500$^{\circ}$C and then put into diluted hydrochloric acid for $\approx 10$\,min. This resulted in plate-like single crystals with a side length of typical 200\,$\mu$m, but in some cases going up to more than one millimeter. 
In parts, the crystals formed large clusters with a common $c$-axis and a mass of up to 40\,mg. X-ray powder diffraction patterns of ground single crystals were recorded on a Stoe diffractometer in transmission mode using Cu-K$_{\alpha}$ radiation. 
The lattice parameters were found to be $a = 4.002(1)$\,\AA\,and $c = 8.647(2)$\,\AA\,and correspond well with the literature data.\cite{quebe2000,chen2008,zhao2008,mcguire2009}
Energy dispersive X-ray (EDX) analysis revealed a stoichiometric Ce:Fe:As content and confirmed the existence of oxygen. In addition, carrier gas-hot extraction (LECO, TCH 600) was used to determine the oxygen content $x_{\rm O}$. The result of $x_{\rm O}  = (23.5 \pm 1.6)$\,at.\% indicates a stoichiometric oxygen occupancy. 
In contrast to the case of Sn-flux-grown 122 single crystals, no Sn peak could be observed in the microprobe spectra of the CeFeAsO single crystals, indicating that the problem of Sn incorporation into the single crystals is much less severe for the 1111 than for the 122.\cite{sun2009} 
The sharpness of the different transitions as well as the high residual resistivity ratio (see below) confirm that Sn (or other crucible elements) incorporation is not a significant issue in these single crystals. 
Specific heat and electrical resistivity measurements were carried out in the temperature range of 1.8 to 300\,K by using a commercial physical property measurement system (PPMS) of Quantum Design. The specific heat was determined by means of a heat-pulse relaxation technique and electrical resistivity was measured in a standard four probe geometry.
The thermal expansion coefficient, $\alpha(\textit{T})=\textit{l}^{-1}(\partial \textit{l}/\partial \textit{T})$, was measured using a high-resolution  capacitive dilatometer built after\,\cite{Pott1983}, which enables relative length changes $\Delta l/l \geq 10^{-10}$ to be resolved.
 
\section{Results}
\begin{figure}
\includegraphics[width=8cm]{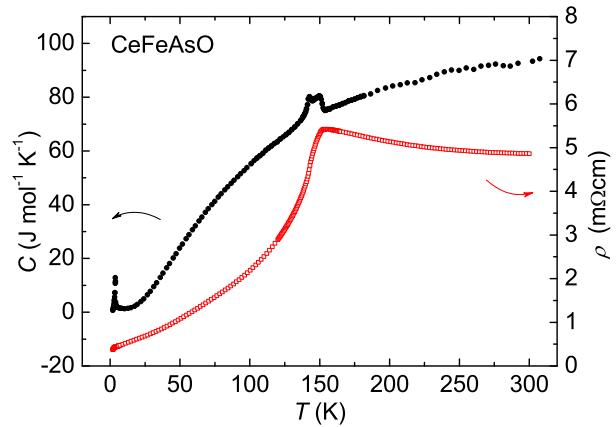}
\caption{\label{hcrhogrob}(color online) Temperature dependence of specific heat (left scale) and electrical resistivity (right scale) of a cluster of CeFeAsO. The resistivity shows an almost step-like drop at $T = 150$\,K and a further decrease towards low temperatures, resulting in RRR = 12. The specific heat shows an anomaly around $T \approx 145$\,K with two distinct maxima, which can be attributed to the structural transition and the AFM ordering of Fe, and a peak at $T_N^{Ce} = 3.7$\,K corresponding to AFM ordering of Ce.\cite{jesche09}}
\end{figure}

\begin{figure}
\includegraphics[width=8cm]{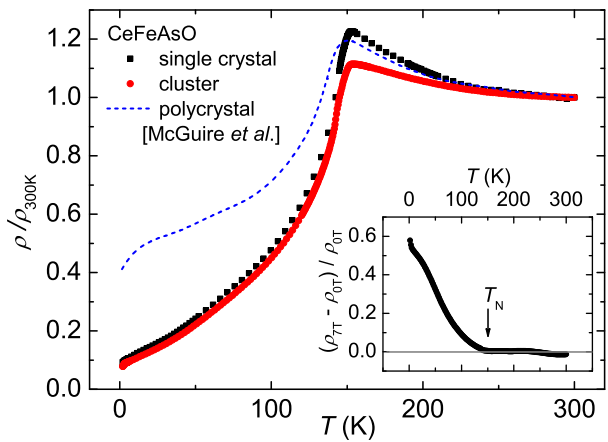}
\caption{\label{rhonormiert+inset}(color online) Temperature dependence of the electrical resistivity normalized to RT values for a small single crystal and a larger cluster of co-aligned single crystals of CeFeAsO. For comparison, the resistivity of polycrystalline CeFeAsO taken from McGuire $et\,al.$\,\cite{mcguire2009} is included. The inset shows the magneto-resistance of single crystalline CeFeAsO with magnetic field applied along the crystallographic $c$-axis and electrical current in the $ab$-plane.}
\end{figure}

For a first overview, we show in Fig.\,\ref{hcrhogrob} the resistivity and the specific heat of a CeFeAsO cluster in the whole investigated temperature range, 1.8 - 300\,K. Below 300 K the resistivity first increases slightly with decreasing temperature. A pronounced drop at 150 K and a further kink at a slightly lower temperature mark the structural transition and the antiferromagnetic (AFM) ordering of Fe, respectively. 
In our single crystals, $\rho(T)$ continues to decrease monotonously with $T$ down to the lowest temperature, resulting in a large RRR $\approx 12$. 
The specific heat $C(T)$ above 20 K is dominated by the phonon contribution, which, at higher temperatures, approaches asymptotically the expected Dulong Petit value $12 R \cong 100$\,J/(mol\,K). 
On top of this smooth contribution, we observe in the $T$-range 140 - 150 K a rather sharp double structured peak connected with the structural transition and the Fe - AFM ordering. 
The peak in $C(T)$ at low temperature is associated with the AFM ordering of Ce at $T_N^{Ce} = 3.7$\,K.\cite{jesche09}
Compared to previously published data, the most prominent features of these results are the sharpness of the transitions and the pronounced metallic behavior resulting in the large RRR. 
This is illustrated in Fig.\,\ref{rhonormiert+inset}, where we compare the resistivity (normalized to the RT value) measured on a small single crystal (electrical current $\parallel ab$), on a larger cluster of (co-aligned) single crystals, and the resistivity of the polycrystal published by McGuire $et~al.$.\cite{mcguire2009} 
The residual resistivities, $\rho_0$, and the values at 300 K increase slightly from the single crystal ($\rho_0 = 140$\,$\mu\Omega$cm and $\rho_{300 \rm K} = 1.6$\,m$\Omega$cm) to the cluster ($\rho_0 = 390$\,$\mu\Omega$cm and $\rho_{300 \rm K} = 4.8$\,m$\Omega$cm), but $\rho_0$ is much larger for the polycrystal ($\rho_0 = 2$\,m$\Omega$cm and $\rho_{300 \rm K} = 4.9$\,m$\Omega$cm). 
All samples show an increase in the resistivity when they are cooled down from RT to $T = 150$\,K, which was also reported for several $R$FeAsO (see e.g.\,\cite{mcguire2009}). This increase is almost the same for the single crystal and for the polycrystal despite quite different resistivities at low temperatures. 
This suggests this increase to be an intrinsic property, pointing either to an increase of the scattering rate or the opening of a gap.
Furthermore, its presence in our single crystal implies that the pronounced metallic behavior, observed for $T < 150$\,K, is not an artifact due to a short-circuit by Sn-flux inclusions. 
Although the increase between RT and 150 K is more pronounced in the single crystal than in the cluster, both curves merge below $T_N$ down to the lowest temperatures.
The inset (Fig.\,\ref{rhonormiert+inset}) shows the transversal magneto-resistance, $\rho_B = (\rho_{7\rm T}-\rho_{0\rm T})/\rho_{0\rm T}$, of a single crystalline sample with magnetic field parallel to the $c$-axis and electrical current in the basal plane. 
At low temperatures, we observe a large positive magnetoresistance of about 60\%, nearly a factor of 3 larger than in the best reported polycrystals \cite{mcguire2009}, a further indication for the higher quality of our single crystals. 
With increasing temperature, $\rho_B$ decreases almost linearly with $T$ and nearly vanishes at $T_N$. 
Both the large positive values observed at the lowest temperatures, where spin disorder scattering is frozen out, and the continuous decrease with increasing $T$ (and increasing $\rho_{0\rm T}$), indicate that  $\rho_B$ is dominated by conventional magnetoresistance effects.

\begin{figure}
\includegraphics[width=8cm]{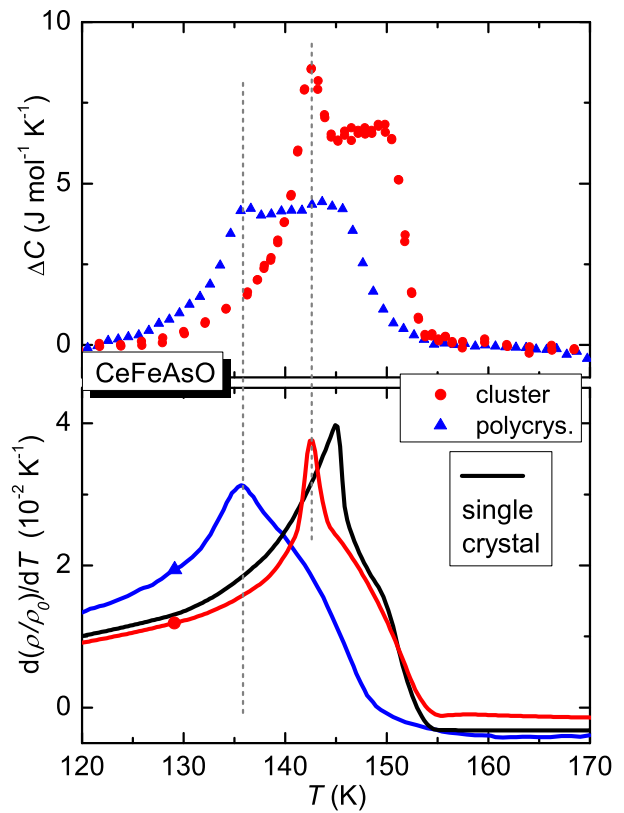}
\caption{\label{rhoAbleitung+HC}(color online) Specific heat (upper panel) and derivative of the electrical resistivity normalized to RT (lower panel) of different CeFeAsO samples. While the onset of the structural transition is close to $T \approx 155$\,K for all samples, the mid point of this transition is slightly shifted to lower temperatures in the polycrystalline sample. In contrast, $T_N$ (peaks with dotted line) decrease significantly from the single crystal, to the cluster and the polycrystal. The data for the polycrystal were taken from \cite{mcguire2009}.}
\end{figure}
The data shown in Fig.\,\ref{rhonormiert+inset} already suggest the transitions to be much sharper in our single crystals than in polycrystalline samples. 
We therefore focus now on the structural transition and the AFM ordering of Fe and compare in Fig.\,\ref{rhoAbleitung+HC} the specific heat as well as the derivative of the resistivity of a small single crystal (only d$\rho(T)/$d$T$), the cluster, and the best data presently available for polycrystalline material (McGuire $et$\,$al.$\,\cite{mcguire2009}) in the temperature range 120 to 170\,K. 
For the specific heat we show only the part related to the transitions. 
For this purpose, a polynomial was fitted to the phononic background well above and well below the transition region and then subtracted from the experimental data in the transition region. 
The low mass ($m = 0.5$\,mg) of the single crystal prevented a precise determination of the part related to the transitions, which amount to less than 0.3\,\% of the total $C(T)$ because of the large phonon background of the sample itself and of the sample holder. 
Thus, for the small single crystal we rely on the d$\rho(T)$/d$T$ data and their relationship to the $\Delta C(T)$ data as established for the cluster and the polycrystal.
The specific heat of the cluster shows a mean-field-type transition at $T_0 = 151$\,K, followed by a sharp peak at a slightly lower temperature $T_N = 142.5$\,K. The analysis of the heating and cooling parts of relaxation curves did not reveal evidences for a thermal arrest and therefore gave no hint for a first order transition, in contrast to the results for the 122 compounds.  
The assignment of the former one to the structural transition and the latter one to the magnetic transition has been well established by many investigations on polycrystals. 
In the polycrystal, the onset of the structural transition seems to be at the same temperature as that for the single crystal, but the transition is much broader, resulting in a lower midpoint $T_0$, and the size of the anomaly ($\Delta C(T)$ value at the plateau) is much smaller. 
The peak at $T_N$ is barely visible, the clear signature for the magnetic phase transition being now the pronounced drop in $\Delta C(T)$ just below $T_N$, which is at a significantly lower $T$ than in the cluster. 
The enthalpy connected with both transitions can be calculated by integrating $\Delta C(T)$ over $T$. We get a larger value for the cluster, $H = 93(5)$\,Jmol$^{-1}$, than for the polycrystal, $H \approx 71$\,Jmol$^{-1}$. 
Nevertheless, the result for the cluster is only half the value $H = 200$\,Jmol$^{-1}$ reported for the combined, first-order magnetic and structural transition in SrFe$_2$As$_2$\,\cite{krellner2008}.

In metallic magnetic systems, the specific heat and the resistivity are related\,\cite{fisher68,krellner07,campoy06}. 
Thus, in the vicinity of magnetic transitions, the derivative of the resistivity often mimics $C(T)$. 
Therefore, we plot in the lower panel of Fig.\,\ref{rhoAbleitung+HC} the derivative of the electrical resistivity (normalized to RT) for the polycrystal, the cluster, and the single crystal. 
The analogy with the $C(T)$ data in the upper part is evident. 
All samples show an asymmetric anomaly with a shoulder at higher temperatures followed by a peak at lower temperatures. 
For the cluster, the positions of the peak in d$\rho(T)/$d$T$ and $C(T)$ match perfectly, while for the polycrystal, the peak in d$\rho(T)/$d$T$ is located at the onset of the drop in $C(T)$. 
Thus, the peak in d$\rho(T)/$d$T$ being more pronounced than the peak in $C(T)$ is an excellent mark for $T_N$. 
In contrast, the anomaly at higher temperatures related to the structural transition is sharper in $C(T)$ than in d$\rho(T)/$d$T$, suggesting that $T_0$ (midpoint) is related with the inflection in the latter quantity. 
Notably, the single crystal not only shows a sharper anomaly, but also a smaller separation between $T_0$ and $T_N$. 
Comparing the three samples (see also Table\,\ref{deltaT} below), $T_0$ stays at 151\,K for the single crystal and the cluster, and shifts to $\approx 147$\,K in the polycrystal, while $T_N$ shifts more strongly from $T_N = 145$\,K in the single crystal to 142.5\,K in the cluster and 136\,K in the polycrystal. 
Thus, the difference $T_0 - T_N$ decreases from about 18\,K in the first study on LaFeAsO\,\cite{delacruz2008} and CeFeAsO\,\cite{zhao2008}, to 11\,K in the polycrystal\,\cite{mcguire2009}, 8.5\,K in the cluster, and to 6\,K in the single crystal.

\begin{figure}
\includegraphics[width=8cm]{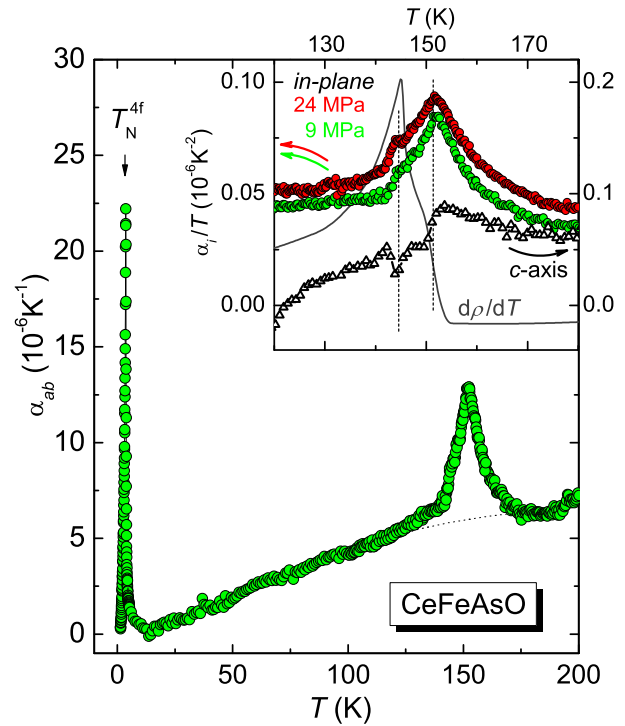}
\caption{\label{thermausdehnung}(color online) Linear thermal expansion coefficient for single crystalline CeFeAsO measured along a non-specified in-plane direction $\alpha_{ab}$.
The dotted line, interpolating between the data outside the peak anomaly, indicates the background expansivity.
The inset shows $\alpha_{ab}$ for two different initial pressure values on the dilatometer cell (left scale) and $\alpha_{c}$ (right scale) in a representation $\alpha_i$/$T$ vs. $T$. The derivative of the electrical resistivity in arbitrary units is shown for comparison. 
The vertical dashed lines indicate the transition temperatures of the structural transition, $T_0 = 151$\,K, and the AFM ordering of Fe moments, $T_{N} = 145$\,K.}
\end{figure}

In the main panel of Fig.\,\ref{thermausdehnung}, we show the thermal expansion coefficient, $\alpha(\textit{T})$, of single crystalline CeFeAsO measured below 200\,K along a non-specified axis within the $ab$-plane, $\alpha_{ab}$.
Upon cooling, a pronounced peak-like anomaly centered at $T = 151$\,K is observed, which is obviously connected with the structural transition. 
The anomaly sits on top of a positive background thermal expansion, cf.\,dotted line in Fig.\,\ref{thermausdehnung}, which smoothly varies with temperature. 
Upon further cooling, $\alpha_{ab}$ decreases monotonously down to $T^{Ce}_N = 3.7$\,K, where a huge $\lambda$-like phase transition anomaly associated with the antiferromagnetic ordering of the Ce-4$f$ moments is visible.\cite{jesche09}
Here, we need to make a remark on the in-plane anisotropy: While in the tetragonal phase above $T_0$ the thermal expansion is isotropic within the basal plane, a strong difference should appear for $T \leq T_0$ between the two orthorhombic in-plane directions. 
At the structural transition itself, one expects huge uniaxial expansion coefficients (on the scale of Fig.\,\ref{thermausdehnung}) of opposite sign along the $a$ and $b$ directions because the orthorhombic distortion $\delta = (a-b)/(a+b)$ increases to $\approx 0.2$\,\% within a small temperature interval\cite{kimber08}.
However, the structural transition ought to result in the formation of twins, and therefore our experiment only probes some mean value, depending on the actual domain structure, of the thermal expansion in the basal plane, which is much smaller.
In fact, results of lattice parameters for the related compound PrFeAsO (Fig. 2b in Ref.\,\cite{kimber08}) suggest that the uniaxial pressure dependence of $T_0$ for pressure applied along the $a$- and $b$-axes are of opposite sign, i.e. according to the Ehrenfest theorem for second-order transitions, $\Delta\alpha_a > 0$ and $\Delta\alpha_b < 0$. Hence, the peculiar form of $\alpha_{ab}$ might be the result of two counteracting effects along $\alpha_a$ and $\alpha_b$, which partly compensate each other in a multi-domain structure.
Attempts to induce a preferential domain orientation by increasing the uniaxial pressure, exerted by the dilatometer cell on the crystal, from about 9\,MPa to 24\,MPa (see inset Fig.\,\ref{thermausdehnung}), failed as they led to practically identical results. 
In the inset of Fig.\,\ref{thermausdehnung} we focus on the $T$ region around the structural and the magnetic transitions and show the details of the in-plane thermal expansion coefficient $\alpha_{ab}$ (left scale), and also the data taken along the $c$-axis, $\alpha_{c}$ (right scale) on an enlarged temperature scale in a representation $\alpha_i$/$T$ vs. $T$. 
Two main observations can be made:

(1) The data are strongly anisotropic both with regard to the anomaly, being peak-like for $\alpha_{ab}$ and jump-like with an additional structure at slightly lower temperatures for $\alpha_{c}$, as well as for the background contribution. 
While $\alpha_{ab}$ is positive in the whole temperature range investigated and, except for the anomaly, decreases smoothly towards low temperatures, cf.\,main panel of Fig.\,\ref{thermausdehnung}, $\alpha_{c}$ is about twice as large as $\alpha_{ab}$ above $T_0$, but then decreases rapidly with decreasing $T$ and even changes sign and becomes negative below $T \simeq 120$\,K. 
This indicates the presence of an anomalous background contribution in this temperature range which causes the $c$-axis to expand upon cooling. 
A similar anisotropy, as well as the presence of a negative thermal expansion along the $c$-axis below 50\,K has been reported for PrFeAsO.\cite{kimber08} 
Furthermore, a similar behavior, with a large $\alpha_c$ (compared to $\alpha_{ab}$) above $T_0$, changing to a very small or negative $\alpha_c$ below $T_0$, has also been reported for undoped BaFe$_2$As$_2$ \cite{budko2009a}, while Co- or Cr-doped BaFe$_2$As$_2$ show a positive $\alpha_c$ at low temperatures (in the non-superconducting state).\cite{budko2009a, budko2009b} 
Thus, a negative or very small positive thermal expansion perpendicular to the FeAs-layers at low temperatures, below the structural transition, changing rapidly to a large value above these transitions seems to be a common feature of the undoped 1111 and 122 FeAs systems, the negative value being related to a well-established antiferromagnetic state. 

(2) The comparison with specific heat and resistivity data allows a more precise analysis of the data close to the transitions. 
The vertical dashed lines at $T_{0} = 151$\,K and $T_{N} = 145$\,K in the inset of Fig.\,\ref{thermausdehnung} indicate the transition temperatures of the structural transition and of the AFM ordering of Fe moments, respectively. 
The former one manifests itself as a sharp peak in $\alpha_{ab}$ and a jump-like discontinuity in $\alpha_{c}$ with $\Delta \alpha_c\mid_{T_0}$ = lim$_{T \rightarrow T_{0}}(\alpha_{T < T_{0}} - \alpha_{T > T_{0}}) \approx -\,3 \times 10^{-6}$\,K$^{-1}$. According to the Ehrenfest relation, this negative discontinuity corresponds to a negative uniaxial-pressure dependence d$T_0$/d$P_c <$ 0. 
In addition, for both directions the anomalies are similar to those observed in Co-doped BaFe$_2$As$_2$.\cite{budko2009a} 
Interestingly, there is a large tail in the thermal expansion data above $T_0$, more pronounced in $\alpha_{ab}$ than in $\alpha_c$, pointing to fluctuations of an order parameter for $T > T_0$.
They are likely not visible in the $C(T)$ data because the ratio between the transition anomaly and the background contribution is much smaller for the specific heat than it is for the thermal expansion. 
Thus, thermal expansion is, in this respect, more sensitive than the specific heat. 
The presence of fluctuations above $T_0$ has already been suggested in thermal expansion measurements performed on polycrystals by Wang $et~al.$\,\cite{wang09} and Klingeler $et~al.$.\cite{klingeler09} 
They proposed these fluctuations to be also responsible for the increase in $\rho(T)$ between RT and $T_0$. However, in $\alpha(T)$ the additional contribution indeed almost diverges towards $T_0$, while the additional contribution in $\rho(T)$ increases only very smoothly towards $T_0$, questioning critical fluctuations as the possible origin for the latter.
 
The signatures at the magnetic transition are very different. In the basal plane we observe at $T_N = 145$\,K a tiny peak followed by a drop in $\alpha_{ab}(T)/T$.
The anomaly along the $c$-axis is much smaller, and the scatter of the data do not allow for a strong statement about its form and size. 
However, the data suggests the presence of a small positive discontinuity $\Delta \alpha_c\mid_{T_{N}} > 0$.

\section{Discussion}
\begin{table}
\caption{Transition temperatures of structural distortion and magnetic ordering of iron for different CeFeAsO samples.}
\begin{center}
	\begin{tabular}{cccc}
	\hline
	\hline
	Sample 							&~~~~$T_0$\,(K)&~~~~$T_N$\,(K)	&~~~~$\Delta T$\,(K)\\
 	\hline
	initial report\,\cite{zhao2008} 	&	158			&  	$\sim 140$	&	$\sim 18$ 		\\
	improved polycrystal\,\cite{mcguire2009} 	&	147			&	136			&	11 	 			\\
	cluster								&	151			&	142.5		&	8.5				\\
	single crystal						&	151			&	145			&	6				\\
	\hline
	\hline
	\end{tabular}
\end{center}
\label{deltaT}
\end{table}
The first important result of the investigation on our single crystals is that the splitting of $\approx 20$\,K between structural transition and magnetic order, reported previously for undoped RFeAsO system, is not an intrinsic property, but can be tuned to much lower values. Table\,\ref{deltaT} shows a summary of measured values of $T_0$, $T_N$, and $\Delta T$. 
The fact that our single crystals present larger and much sharper anomalies at these transitions, as well as a much higher residual resistivity ratio RRR, $\approx 12$, indicate that they are of better quality, i.e. with less defects than the previously reported single crystals or polycrystalline materials. 
Thus, the reduction of the splitting $\Delta T = T_0 - T_N$ is seemingly related to an increase of the sample quality, i.e.\,a decrease of the amount of defects. 
Interestingly, it was recently shown that while in pure BaFe$_2$As$_2$ the structural and the magnetic transition are intimately connected ($\Delta T = 0$), doping by partial substitution of Co on the Fe site results in a splitting of the two transitions, with $\Delta T$ increasing with Co content to $\approx 15$\,K at 5\% Co substitution.\cite{ni08,chu09,lester09,pratt09} 
Our results demonstrate an analogous phenomenon in the RFeAsO compounds, though here the large splitting initially reported was likely due to (unintentionally) imperfect samples. 
Thus, while splitted or common structural/magnetic transitions were thought to be specific properties of 1111 and 122 compounds, respectively, the results on Co doped BaFe$_2$As$_2$ and on our single crystals indicate that also in this aspect both type of compounds are very similar.
The question is now, which mechanism leads to this splitting. 
One proposition, which emerged from a localized moment approach, is that the splitting is a consequence of a strong two-dimensional character, i.e. of a very weak interlayer magnetic exchange.\cite{fang08,xu08,qi09}
It is based on older theoretical studies of the frustrated square lattice. 
Long before the discovery of the superconductivity in the Fe-pnictides, P. Chandra $et\,al.$ proposed for this model the occurrence of an Ising phase transition preceding the transition to the magnetically ordered state.\cite{chandra90} 
This Ising order-parameter is related to the relative orientation of two weakly coupled sublattices with (fluctuating) N$\acute e$el magnetization $m_1$ and $m_2$, corresponding to the two Fe atoms occupying one unit cell: $\phi(x) = m_1(x) \cdot m_2(x)$, where $x$ stands jointly for space and time variation.\cite{fernandes09} 
This scalar product does not change sign upon magnetic field inversion and hence corresponds to a nematic phase. 
Its transition temperature depends on the intra-plane exchange, but not on the inter-plane exchange. 
In contrast, for an isotropic Heisenberg magnetic moment (which is a good approximation for the present case), the transition temperature for true magnetic long-range order decreases with inter-plane exchange, down to $T_N = 0$ for a pure two-dimensional case. 
Therefore, increasing the two-dimensional character will first shift $T_N$ below $T_0$ and then increase the separation. There is indeed evidence that the two-dimensional character is more pronounced in the 1111 than in the 122 compounds.\cite{qi09} 
Furthermore, the effective coupling between adjacent FeAs planes in the 1111 seems to be frustrated,~i.e.~on the verge between ferromagnetic and antiferromagnetic, since the Fe-moments in LaFeAsO and NdFeAsO order antiferromagnetically along the $c$ direction, while they are ferromagnetically aligned in CeFeAsO and PrFeAsO.\cite{zhao2008} 
In the 122 family one always observes antiferromagnetic ordering along $c$. 
Thus, the model with the nematic phase was in nice agreement with the initial picture of a significant splitting for the 1111 but no splitting for the 122. 
However, the results on the Co-doped BaFe$_2$As$_2$ single crystals and on our CeFeAsO single crystals raise some question marks, since they put the 1111 and the 122 system closer to each other, much closer than expected from the difference in the inter-plane exchange.\cite{qi09} 
The observation of an increase of the splitting upon increasing Co substitution or increasing defect concentration indicates that predominantly defects are at the origin of this splitting, and not a strong difference in the inter-layer exchange. 
In this context, one should also note that the closeness of $T_N$ in undoped BaFe$_2$As$_2$ ($T_N \approx 145$\,K) and RFeAsO compounds $(T_N \approx 135$\,K) is (within a simple $J_1 - J_2 - J_z$ model) in clear disagreement with a significant difference in $J_z$ between these two types of compounds. 
On the one hand, in order to keep the nematic-phase model, one would now have to argue that disorder decreases the coherence along the $c$-directions and thus the effective coupling between adjacent planes. 
On the other hand, in our comparison between different samples we see that the onset of the structural transition stays remarkably stable at 155\,K, while $T_N$ shifts significantly, which is exactly the behavior expected for the nematic phase model. 
Thus, our experimental results give arguments both in favor and against the nematic phase model.

The sharper anomalies observed in our single crystals allow also for a more precise discussion of their nature: At $T_0$ we observe a clear mean-field type anomaly in $C(T)$, a clear step in $\alpha_c$, and a progressive change of slope in $\rho(T)$, which indicate this transition to be second-order. 
At $T_N$ we observe a sharp peak in d$\rho(T)$/d$T$ and in $C(T)$, but only small anomalies in $\alpha(T)$. The two former features indicate either a sharp, $\lambda$-type second-order or a broaden first-order type transition. 
For the nematic phase model, both transitions were suggested to be second-order.\cite{qi09} 
Furthermore, we note that the anomalies we observed in our CeFeAsO single crystals in $C(T)$, in $\alpha(T)$ (including the anisotropy), and in d$\rho(T)/$d$T$ (there the sign has to be inverted) at $T_0$ and at $T_N$ are very similar to those reported in Co-doped BaFe$_2$As$_2$\,\cite{chu09,budko2009a}, stressing again the similarity between the RFeAsO and the AFe$_2$As$_2$ type of compounds.

\section{Conclusions}
In summary, we adapted a high-temperature Sn-flux technique to the growth of RFeAsO compounds and obtained comparatively large CeFeAsO single crystals. 
Investigation of their properties revealed much sharper anomalies at the structural and magnetic phase transitions as well as a much lower residual resistivity and higher residual resistivity ratio than reported previously for polycrystalline samples or single crystals. 
This demonstrates the comparatively high quality of our single crystals. They present a pronounced and continuous  decrease of the resistivity below the structural and magnetic transitions, leading to a RRR of $\approx 12$. 
This proves a well-defined metallic character for transport within the FeAs plane in the antiferromagnetically ordered state, and excludes a Mott metal-insulator transition.
The mean-field type anomalies observed at $T_0 \cong 151$\,K in $C(T)$ and $\alpha_c$ indicate the structural transition to be of second-order type. 
In contrast, sharp peaks in $C(T)$ and d$\rho(T)/$d$T$ at $T_N = 145$\,K are compatible with either a sharp $\lambda$-type transition or a broaden first-order type transition. 
These anomalies are very similar to those observed in Co-doped BaFe$_2$As$_2$, where the Co-doping leads to a splitting of the structural and antiferromagnetic transitions. 
A comparison of different samples, including previously published data, reveals a decrease of the splitting between structural and AFM transition from $\approx 18$\,K in the early reports on CeFeAsO, to 11\,K in the best reported polycrystals, 8.5\,K in a cluster of co-aligned crystals, and 6\,K in a single crystal. 
Thus, this splitting is not an intrinsic property of CeFeAsO (or other RFeAsO) but can be tuned by the sample preparation conditions. 
Our results demonstrate that a better sample quality results in a smaller splitting, indicating that the splitting is at least partially induced by defects. 
This is supported by the appearance of a splitting upon Co-doping BaFe$_2$As$_2$. Whether this splitting shall disappear in a perfect RFeAsO sample, as for pure AFe$_2$As$_2$ compounds, remains an open question. 
Our results, in connection with the published results on Co-doped BaFe$_2$As$_2$, indicate that the coupling between magnetic and structural transitions is very similar in both type of compounds, much more similar than previously thought. 
This questions the idea of an Ising nematic order parameter proposed to explain the splitting and the differences between 1111 and 122 compounds. 
However, the fact that the decrease of the splitting results from a shift of $T_N$ to lower $T$ with decreasing sample quality, while $T_0$ does not change, is in agreement with this nematic phase concept. 

\section{Acknowledgment}
R. Weise is acknowledged for technical support. The authors thank P. Scheppan, U. Burkhardt, and G. Auffermann for chemical analysis of the samples. This work was performed within the framework of the SPP 1458 priority program of the Deutsche Forschungsgemeinschaft.

\end{document}